\documentclass[fleqn,usenatbib]{mnras}

\usepackage{newtxtext,newtxmath}

\usepackage[T1]{fontenc}
\usepackage{ae,aecompl}

\hypersetup{draft}

\usepackage{graphicx}	
\newcommand{\src}{BW Cir}
\newcommand{\srctwo}{GS~1124-684}
\newcommand{\xrb}{BHXB}
\newcommand{\nh}{N_{\rm H}}   
\newcommand{\fr}{f_{\rm 5\,GHz}}
\newcommand{\fx}{F_{\rm 1-10\,keV}}
\newcommand{\lr}{L_R}
\newcommand{\lx}{L_X}
\newcommand{\flux}{{\rm erg}\,{\rm s}^{-1}\,{\rm cm}^{-2}}
\newcommand{\ergs}{{\rm erg}\,{\rm s}^{-1}}
\newcommand{\mbh}{M_{\rm BH}}
\newcommand{\msun}{M_\odot}

\newcommand{\srcgaia}{7.1^{+4.8}_{-3.9}}   
\newcommand{\srctwogaia}{5.4^{+3.7}_{-2.4}}  

\defcitealias{gandhi19}{G19}
\defcitealias{casares04}{C04}
\defcitealias{casares09}{C09}

\title[Quiescent Radio Jets]{Toward a Larger Sample of Radio Jets from Quiescent Black Hole X-ray Binaries}

\author[R. M. Plotkin et al.]{
R. M. Plotkin,$^{1}$\thanks{E-mail: rplotkin@unr.edu}
A. Bahramian,$^{2}$
J. C. A. Miller-Jones,$^{2}$
M. T. Reynolds,$^{3}$
P. Atri,$^{4}$
\newauthor 
T. J. Maccarone,$^{5}$
A. W. Shaw,$^{1}$
and P. Gandhi$^{6}$
\\
$^{1}$Department of Physics, University of Nevada, Reno, NV 89557, USA\\
$^{2}$International Centre for Radio Astronomy Research, Curtin University, GPO Box U1987, Perth, WA 6845, Australia\\
$^{3}$Department of Astronomy, University of Michigan, 500 Church Street, Ann Arbor, MI 48109, USA\\
$^{4}$ASTRON, Netherlands Institute for Radio Astronomy, Oude Hoogeveensedijk 4, 7991 PD Dwingeloo, The Netherlands\\
$^{5}$Department of Physics, Box 41051, Science Building, Texas Tech University, Lubbock, TX 79409-1051, USA\\
$^{6}$Department of Physics and Astronomy, University of Southampton, Highfield, Southampton SO17 1BJ
}

\date{Accepted XXX. Received YYY; in original form ZZZ}

\pubyear{2021}

\begin{document}
\label{firstpage}
\pagerange{\pageref{firstpage}--\pageref{lastpage}}
\maketitle

\begin{abstract}
Quiescent black hole X-ray binaries (X-ray luminosities $\lesssim 10^{34}~\ergs$) are believed to be fed by  hot accretion flows that launch compact, relativistic  jets.  However, due to their low luminosities, quiescent jets have  been detected in the radio waveband from only five systems so far.  Here, we present  radio observations of two quiescent black hole X-ray binaries with the Australia Telescope Compact Array.  One system, \srctwo, was not detected.  The other system, \src,  was  detected over two different epochs in 2018 and 2020, for which we also obtained quasi-simultaneous X-ray detections with \textit{Chandra} and \textit{Swift}.  \src\ is now the sixth quiescent X-ray binary with a confirmed radio jet.  However, the distance to \src\ is uncertain, and we find that \src\  shows  different behaviour in the radio/X-ray luminosity plane depending on the correct distance.  Estimates based on its  G-type subgiant donor star place \src\ at $>$25 kpc, while initial optical astrometric measurements from \textit{Gaia} Data Release 2 suggested likely distances of just a few kpc. Here, we use the most recent  measurements from \textit{Gaia} Early Data Release 3 and find a distance $d=7.1^{+4.8}_{-3.9}$ kpc and a potential kick velocity PKV=$165^{+81}_{-17}$ km s$^{-1}$, with distances up to $\approx$20 kpc possible based on its parallax and proper motion.  Even though  there is now less tension between the  parallax and donor-star based distance measurements, it remains an unresolved matter, and we conclude with suggestions on how to reconcile the two measurements.

\end{abstract}




\section{Introduction}

Transient black hole X-ray binaries (\xrb s) spend the majority of their  time in a quiescent accretion state, which normally corresponds to X-ray luminosities ($\lx$) between $10^{30} - 10^{34}~\ergs$  \citep[e.g.,][]{corbel06, plotkin13}.  Extrapolating  the `hard' X-ray state  to quiescence (see \citealt{remillard06} for state definitions), the expectation is that quiescent \xrb s are fed by some flavor of a radiatively inefficient accretion flow (RIAF; e.g., \citealt{ichimaru77, narayan95, blandford99, yuan14}).  Quiescent \xrb s are also expected to launch compact, partially self-absorbed synchrotron jets \citep{blandford79, fender01, corbel02}.  However, the radiative power of quiescent jets  is so low that the current number of radio-detected quiescent outflows is limited to only five systems \citep{gallo05, gallo06, gallo14, hynes09, miller-jones11, dzib15, ribo17, corbel13, rodriguez20, tremou20}, out of $\approx$20 dynamically confirmed \xrb s in our Galaxy and at least three times as many candidates \citep{corral-santana16, tetarenko16a}.

Radio detections of quiescent \xrb\ jets are important for self-consistently modeling the higher-energy radiation outputted by  quiescent systems \citep[e.g.,][]{mcclintock03, yuan05, malzac14, markoff15, plotkin15, connors17, moscibrodzka19}. Constraints on  quiescent radio jet variability   are also starting to emerge.  When taken in a multiwavelength context, the radio time domain has the potential to improve our understanding of the structure and energetics of quiescent jets, and how quiescent jets couple to the underlying RIAF \citep[e.g.,][]{ miller-jones08, hynes09, dzib15, rana16, dincer18, gallo19, plotkin19}.  However, we still require a larger sample of radio-detected quiescent \xrb s, or more meaningful limits from  non-detections.  Furthermore, radio constraints should span a range of \xrb\ properties (e.g., orbital period, black hole/donor mass, inclination, etc.), so that we can start to tease out jet and RIAF properties that are common across the entire population vs.\ specific to individual sources.

In addition to advancing our physical understanding, empirical limits on more  quiescent radio jets are crucial for guiding future \xrb\ surveys that rely on radiative signatures to discover new sources.  The `traditional' method of discovering \xrb s during outbursts leads to \xrb\ samples that are unlikely representative of the underlying population in terms of properties like orbital period, inclination, black hole mass, etc. \citep[see, e.g., ][]{narayan05, jonker14}.  
Discovering new systems in quiescence (instead of during outburst) can alleviate some selection biases, and techniques have already been suggested/implemented in the X-ray \citep{agol02, jonker14}, optical \citep{casares18}, and radio wavebands \citep{maccarone05, strader12}. Note, these selection  techniques in quiescence also have the potential to reveal isolated black holes. 

Radio selection of accreting \xrb s  has a unique advantage over  other wavebands, in that  radio emission is not affected by dust obscuration or crowding by other sources in the field.  Indeed, radio surveys are  already revealing  new candidate \xrb\ populations, including black hole candidates in Galactic globular clusters \citep{strader12, chomiuk13, miller-jones15, shishkovsky18, zhao20}.     The most common radio diagnostic so far (at least for initial candidate identification) has been to compare the ratio of radio luminosity ($\lr$) to X-ray luminosity, which increases for decreasing $\lx$ according to the  radio/X-ray luminosity correlation for hard state and quiescent \xrb s  \citep[e.g.,][]{corbel13, gallo18}.  To efficiently identify new quiescent systems, one still desires a larger comparison sample of dynamically-confirmed quiescent \xrb s with radio detections, so that one can confidently parameterise the multiwavelength luminosity  space that is occupied by \xrb s. 

The second data release (DR2) of the \textit{Gaia} mission \citep{gaiamission, gaia-collaboration18} included  trigonometric parallax measurements to 11 \xrb s (\citealt{gandhi19}, hereafter \citetalias{gandhi19}).  The new \textit{Gaia} distance constraints motivated us to reassess if any dynamically-confirmed \xrb s might be close enough to present a realistic chance of a radio detection.  We stress, however, that the majority of parallaxes were at the $\lesssim$2--3$\sigma$ level.  At this level, translating the parallaxes to distances is best achieved via Bayesian modeling, which requires using a prior that describes how \xrb s are distributed in our Galaxy (e.g., \citealt{grimm02, atri19}; \citetalias{gandhi19}).  In turn, implied \textit{Gaia} parallax-based distances can be highly dependent on the adopted prior, and most distances are less precise than distances derived from other methods.  See \citet{atri19} and \citetalias{gandhi19}   for comparisons on distances inferred from  \textit{Gaia} parallaxes vs. other methods (\citealt{jonker04} provide a summary of other  distance estimation techniques).

 Despite the above caveats, we identified two dynamically confirmed \xrb s in \textit{Gaia} DR2 worth  radio scrutiny: \src\ and \srctwo. In Section~\ref{sec:targs} we describe properties of each target.  In Section~\ref{sec:obs} we present new observations of each source in quiescence, which included radio observations with the Australia Telescope Compact Array (ATCA) for both sources, and also quasi-simultaenous X-ray observations with the \textit{Chandra X-ray Observatory} and the \textit{Neil Gehrels Swift Observatory} for \src.  Results are presented in Section~\ref{sec:res}, which are then discussed in Section~\ref{sec:disc}.  We conclude with suggestions for future work in Section~\ref{sec:disc:future} and a summary in Section~\ref{sec:summary}.  Throughout, all errors are reported at the 68\% confidence level, unless stated otherwise.  Although our initial target selection was based on \textit{Gaia} DR2, all astrometric information in this paper is taken from \textit{Gaia's} Early Data Release 3 \citep[EDR3;][]{gaiaedr3}, which was released after our observations were taken.

\section{Targets}
\label{sec:targs}

\begin{figure*}
	\includegraphics[width=15cm]{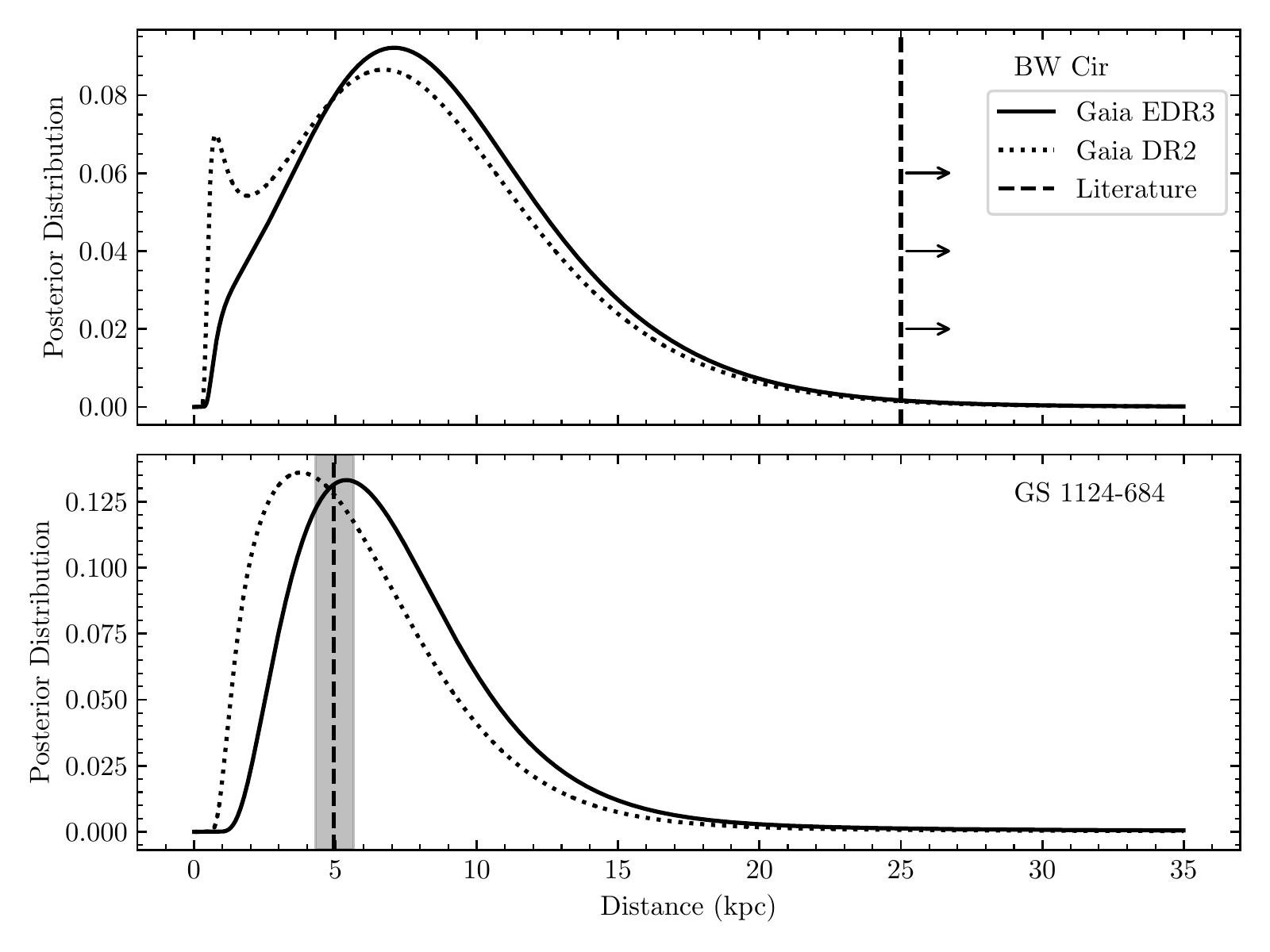}
    \caption{Distance posterior distributions from \textit{Gaia} for \src\ (top panel) and \srctwo\ (bottom panel), following the methodology from \citet{atri19}.   The solid black lines show the posterior from the  \textit{Gaia} EDR3 parallax measurement used in this paper, and the dotted lines the posterior from \textit{Gaia} DR2.  The dashed black vertical lines show the distances to each source in the pre-\textit{Gaia} literature based on donor star properties: $>$25 kpc for \src\ (\citetalias{casares09}) and $4.95^{+0.69}_{-0.65}$ kpc  for \srctwo\ \citep{wu16}.  The gray shaded region in the bottom panel denotes the 1$\sigma$ uncertainty on the literature distance. }
    \label{fig:dist}
\end{figure*}

For both of our targets, we derive distance posterior distributions from the EDR3 parallax measurements by applying the Bayesian method developed by \citetalias{gandhi19} and \citet{atri19}.  They adopt a prior based on the three-dimensional distribution of X-ray binaries in our Galaxy by defining their density in the disc, bulge and spheroid of the Galaxy \citep{grimm02}. The EDR3 posterior distributions imply distances of $d=\srcgaia$ and $\srctwogaia$ kpc for \src\ and \srctwo, respectively (Figure~\ref{fig:dist}), where the distances and uncertainties represent the modes and  68\% confidence intervals of the posterior distributions.
The distance estimates in the literature for each source (i.e., prior to \textit{Gaia}) are  $d>25$ kpc for \src\ (\citealt{casares04, casares09}; hereafter \citetalias{casares04}; \citetalias{casares09}), and $d=4.95^{+0.69}_{-0.65}$ kpc for \srctwo\ \citep{wu16}. 
The \textit{Gaia} astrometry and inferred distances are summarized in Table~\ref{tab:gaia}, where we also include the older DR2 measurements for comparison (since our  target selection was initially based on DR2).  
Each target is described in more detail below.

\begin{table*}
	\centering
	\caption{Comparison of \textit{Gaia} EDR3 to DR2 measurements. The reported parallax measurements have been corrected for the zero point offset as described by \citet{luri18} for \textit{Gaia} DR2 and \citet{lindegren20} for \textit{Gaia} EDR3. The distances $d_{\rm post}$ represent the mode and 68\% confidence intervals from the posterior distributions.}
	\label{tab:gaia} 
	\begin{tabular}{cccccccccc}
		\hline
		&
		\multicolumn{4}{c}{\textit{Gaia} EDR3} & 
		\multicolumn{4}{c}{\textit{Gaia} DR2} &
		Lit.\\ 
		Source & 
		$\mu_\alpha \cos \delta$ & 
		$\mu_\delta$ & 
		$\pi$ & 
		$d_{\rm post}$ & 
		$\mu_\alpha \cos \delta$ & 
		$\mu_\delta$ & 
		$\pi$ & 
		$d_{\rm post}$ & 
		$d_{\rm lit}$ \\  
		  & 
		(mas yr$^{-1}$)  & 
		(mas yr$^{-1}$)  & 
		(mas)            & 
		(kpc)            & 
		(mas yr$^{-1}$)  & 
		(mas yr$^{-1}$)  & 
		(mas)            & 
		(kpc)            & 
		(kpc)            \\ 
		\hline
        BW Cir    & $-5.07\pm0.63$  &  $-2.10\pm0.58$ & $1.28\pm0.53$ & $7.1^{+4.8}_{-3.9}$  & $-9.38\pm2.22$ & $-5.70\pm2.26$ & $1.86\pm0.58$ & $6.7^{+4.9}_{-4.6}$ & $>$25$^a$ \\  
        GS 1124$-$64  & $-2.93\pm0.24$   & $-1.39\pm0.26$   &  $0.20\pm0.24$   &  $5.4^{+3.7}_{-2.4}$    &  $-2.44\pm0.61$  &  $-0.71\pm0.46$  & $0.64 \pm 0.34$  & $3.8^{+3.7}_{-2.1}$ & $4.95^{+0.69,b}_{-0.65}$\\
		\hline
	\end{tabular}

	$^a$ Distance based on the donor star from \citetalias{casares09}.
	
	$^b$ Distance based on the donor star from \citet{wu16}
	
\end{table*}

\subsection{\src}
\label{sec:targs:bwcir}

\src\ is a dynamically confirmed black hole, with an orbital period $P_{\rm orb}=2.54451 \pm  0.00008$ d ($\approx$61 h) and a  mass function $f\left(\mbh \right)=5.73\pm0.29 \msun$ \citepalias{casares04,casares09}.   Since the 1960s there have been five X-ray outbursts detected from a region consistent with \src\ (see the summary in \citealt{tetarenko16a}), but only three have been reliably identified as the same X-ray source: outbursts in 1987 \citep{makino87}, in 1997 \citep{remillard97}, and in 2015 \citep{miller15}.

The optical counterpart shows a relatively large amount of stochastic variability in quiescence (a few tenths of a magnitude, depending on the filter; \citetalias{casares09}).  Even with that level of variability, the optical properties of the donor star have been well constrained by \citetalias{casares04} and \citetalias{casares09}: they classify the donor  as a G0-5 III star (surface temperature 5091 -- 5743 K), they detect ellipsoidal modulations (indicating that accretion  is fed by Roche lobe overflow), and they constrain the donor-to-black hole mass ratio to $q=0.12\pm0.04$.  For an inclination $\lesssim 79^\circ$ (based on the lack of X-ray dips), the orbital parameters imply a black hole mass $\mbh \gtrsim 7.6 \pm 0.7 \msun$ and a donor mass $M_d \gtrsim 0.9 \pm 0.3 \msun$.  Stellar evolution arguments for a G sub-giant limit the donor star to $\lesssim$2.4 $\msun$ \citepalias{casares04}, meaning that the donor  is constrained to a relatively narrow range in mass.  Comparing the absolute magnitude of the donor  (which they calculate via the Stefan-Boltzmann law for a  5000 K blackbody radiator filling the Roche lobe of a 1.0 $\msun$ star, which is $\approx3.6~R_\odot$)  to the optical apparent magnitude ($R=20.65$), and accounting for 56\% disk veiling and a reddening of $E\left(B-V\right) \sim 1$ \citep{kitamoto90}, \citetalias{casares09} derive a distance $>$25 kpc. They also determine an upper limit  $<$61 kpc if its  outbursts were Eddington-limited.

Surprisingly, \textit{Gaia} measured a parallax to this source that implies a closer distance.  The latest EDR3 \textit{Gaia} parallax is $\pi = 1.28 \pm 0.53$ milli-arcsec, which corresponds to $\srcgaia$ kpc when adopting a prior based on the distribution of low-mass X-ray binaries in our galaxy \citep{grimm02, atri19}.  We note that the \textit{Gaia} proper motion and parallax measurements all decreased by $1-2\sigma$ in EDR3 compared to DR2, and we suspect that some of these changes are driven by systematics (i.e., they are not caused only by statistical fluctuations related to the longer time baseline of observations in EDR3).  The most obvious warning sign is that the distance posterior distribution  changed from being bimodal in DR2 (peaks near 0.7 and 6.7 kpc) to unimodal in EDR3 (peak at 7.1 kpc; Figure \ref{fig:dist}).  We discuss in \S \ref{sec:disc:gaiaconfusion} that an interloping source could be blended with the \textit{Gaia} optical counterpart, which may explain some of these systematics and help ease tension between the \textit{Gaia} and literature distance estimates.

Possible systematic issues aside, either a small distance (as implied by \textit{Gaia)} or a large distance (as implied by the donor star) has unusual consequences.  We refer the reader to \citetalias{gandhi19} for a detailed discussion based on the DR2 data.  To summarize their discussion,  the large $>$25 kpc distance causes the quiescent X-ray luminosity to be $>$10 times larger than expected from its orbital period \citep{reynolds11}, and based on the DR2 proper motions it also implies a large peculiar velocity $\gtrsim$900 km s$^{-1}$ that starts to approach  the escape velocity of the Milky Way \citepalias{gandhi19}.  This distance  also implies that during its 2015 outburst, \src\ peaked at an  X-ray luminosity in the hard state that was 5-6 times higher than previously observed for any other hard state system \citep{koljonen16, tetarenko16a}.   However, a smaller distance introduces new problems, in that it would require a significantly smaller radius (and therefore mass) for the donor star.  Note, the 25 kpc distance limit already accounts for the lowest donor mass allowed by the mass function and $q$ values (i.e., $\approx 1 \msun$).

 The evidence and assumptions by \citetalias{casares04} and \citetalias{casares09} that point to $>$25 kpc (based on the donor star) are very  robust.  The biggest source of uncertainty is  the level of extinction.  However, tweaking the adopted reddening of $E\left(B-V\right) \sim 1$ within reasonable values (while remaining consistent with the level of X-ray absorption) can only move the distance down to $\approx$16 kpc  \citep{reynolds11}.  Requiring the donor mass to be $>$0.1~$\msun$, and assuming the donor fills its Roche lobe, further provides a strict limit of $d \gtrsim 13$ kpc (see, e.g., \citetalias{gandhi19}). 
 Compared to \textit{Gaia} DR2, the smaller proper motions and parallax measurements in EDR3  now allow a more reasonable probability of the high-distance tail of the posterior distribution extending out to $\approx$20 kpc (with reasonable space velocities; see \S \ref{sec:disc:kindoffar}), but reconciling the \textit{Gaia} and literature-based distances still remains a challenge.

\subsection{\srctwo}
\label{sec:targs:srctwo}

 \srctwo\ (also known as Nova Muscae 1991) is an important source in terms of our understanding of black hole accretion, as it is one of the most high profile systems to which advection dominated accretion flow (ADAF; e.g., \citealt{narayan94, narayan95a}) models were initially tested against \citep{esin97}.  \srctwo\ was discovered during a bright outburst in 1991 \citep{dellavalle91, lund91} and has been in quiescence since the end of that outburst.  It contains a dynamically confirmed black hole with $\mbh = 11.0_{-1.4}^{+2.1}~\msun$ \citep{wu16} in a $0.43260249(9)$ d orbit \citep[i.e., 10.4 h;][]{wu15}, and the system has an inclination angle of $43.2^{+2.1}_{-2.7}$ degrees \citep{wu16}.   The EDR3 \textit{Gaia} parallax ($0.20\pm0.24$ milli-arcsec) implies a distance of $\srctwogaia$ kpc, which is consistent with the most recent pre-\textit{Gaia} distance of $4.95^{+0.69}_{-0.65}$ kpc (which was derived from the properties of a donor star calculated to have a mass of $0.89^{+0.18}_{-0.11} \msun$, a radius $1.06^{+0.07}_{-0.04}~R_\odot$, and a temperature $4400\pm100$ K; \citealt{wu16}). Note, the \textit{Gaia} distance is less precise than the distance inferred from the donor star.

\section{Observations}
\label{sec:obs}
\subsection{Radio observations }
\label{sec:obs:radio}

ATCA observations were taken for both sources through program C3280 (PI Plotkin), and all  data  were obtained over 2$\times$2048 MHz basebands centred at 5.5 and 9.0 GHz.  Details specific to each source are described below.

\subsubsection{\src: 2018 and 2020 with ATCA}

We initially observed \src\ over a 12 hour ATCA run from  2018 October 8 UT 20:40 $-$ 2018 October 9 UT 08:40 in the 6A configuration, yielding $\approx$600 min on source (prior to flagging).  After obtaining a tentative detection (see below), we took deeper observations  on 2020 March 13 UT 08:50 - 20:50 and 2020 March 14 UT 08:50 - 20:50 in the 6D configuration,  obtaining an additional $\approx$1200 min integration.   Data were calibrated following standard procedures in {\tt Miriad} v20190411 \citep{sault95}.  We used 1934$-$638 as the bandpass and flux calibrator, and we cycled to the phase calibrator 1352$-$63 every 10-15 min to solve for the complex gain solutions.  Weather conditions were good during all observations.  

Imaging was performed within the  Common Astronomy Software Applications ({\tt CASA}) v5.6.2  \citep{mcmullin07} using the task {\tt tclean}.  We used two Taylor terms ({\tt nterms=2}) to account for the wide fractional bandwidth, and we used Briggs weighting with {\tt robust=0.5} as a compromise between reducing sidelobes from nearby sources while retaining sensitivity.  For each night we produced images at 5.5 and 9.0 GHz separately, and we also stacked both frequencies to produce images at a central frequency of 7.25 GHz.  For the 2020 observations, we also stacked observations from both nights. Point-like radio emission was detected in both 2018 and 2020 at a right ascension of 13$^h$58$^m$09.73$^s$(0.01) and a declination $-64^\circ 44 \arcmin 05.35\arcsec$(0.08), which is consistent with the optical position from \textit{Gaia}.

\begin{figure*}
    	\includegraphics[width=18cm]{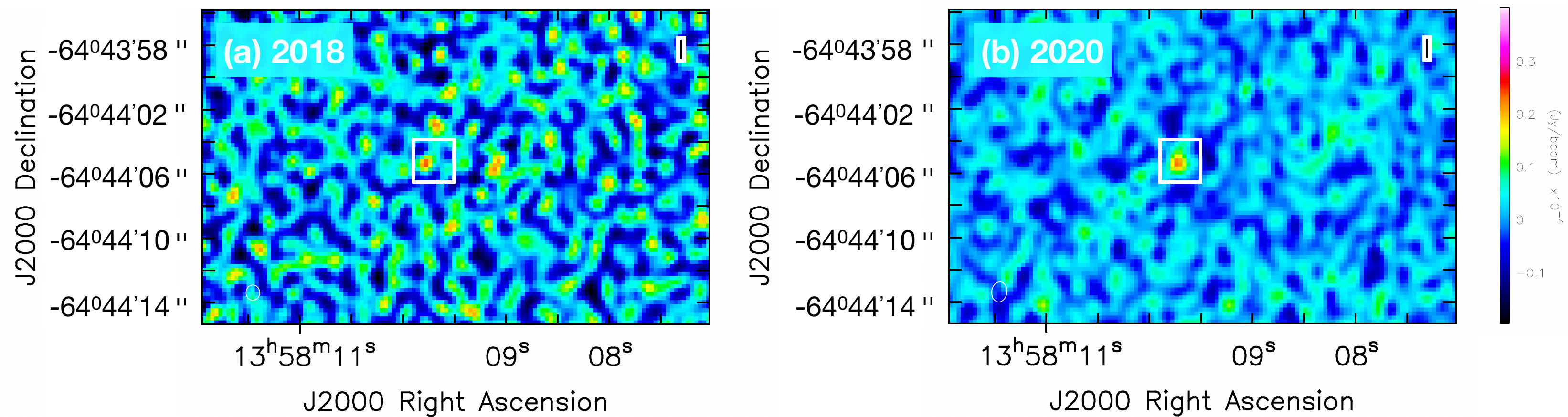}
    \caption{ATCA radio images of \src\ from 2018 (panel a; $\sigma_{\rm rms}=6.4~\mu$Jy bm$^{-1}$ with $\approx 600$ min on source) and 2020 (panel b; $\sigma_{\rm rms}=3.6~\mu$Jy bm$^{-1}$ with $\approx$1200 min on source).  Both images were produced using Briggs weighting (robust=0.5 in {\tt CASA}) at a central frequency of 7.25 GHz after stacking the 5.5 and 9.0 GHz basebands, and using a phase centre at  right ascension 13$^h$58$^m$09.70$^s$ and declination $-64^\circ 44 \arcmin 15.20\arcsec$.  The colors in each image are displayed on a linear scaling (the colorbar on the right side applies to both images).  The white squares, which are 2.5 arcsec on each side, are centred on the optical \textit{Gaia} position of \src.  The source was detected at a similar flux density on both epochs ($21.7\pm6.4\, \mu$Jy in 2018 and $23.6 \pm 3.6\, \mu$Jy in 2020), although the 2020 detection is more significant because of the longer integration time.
    }
    \label{fig:atca}
\end{figure*}

Flux densities were measured within {\tt CASA} using the task {\tt imfit}, using a point source to model the flux density.  We detected radio emission in both 2018 (3.4$\sigma$) and 2020 (6.6$\sigma$).  The radio detections were not strong enough to obtain useful spectral information between 5.5 and 9.0 GHz, so we only report  flux densities after stacking the full $\approx$4 GHz bandwidth (corresponding to a central frequency of 7.25 GHz).  We also do not detect any variability (within the uncertainties) between our two 2020 epochs, so we report only a single flux density from 2020 after combining both days of observations.  We find $f_{7.25\, {\rm GHz}} = 21.7 \pm 6.4$ $\mu$Jy bm$^{-1}$ in 2018 and $f_{7.25\, {\rm GHz}} = 23.6 \pm 3.6$ $\mu$Jy bm$^{-1}$ in 2020, where the uncertainties represent the root mean square (rms) noise measured in a blank area of the sky.\footnote{In 2020, our final rms uncertainties are consistent with expectations from the ATCA online sensitivity calculator.  In 2018, we reached a sensitivity that was $1-2 \mu$Jy higher than  expected, probably because we flagged more data in 2018 due to mildly less stable atmospheric conditions and phase solutions.}  
Flux densities are summarized in Table~\ref{tab:fluxes}, and radio images are presented in Figure~\ref{fig:atca}.

\begin{table*}
	\centering
	\caption{Quiescent fluxes of \src\ and \srctwo}
	\label{tab:fluxes} 
	\begin{tabular}{ccccccccc} 
		\hline
		&
		\multicolumn{4}{c}{Radio} & 
		\multicolumn{4}{c}{X-ray} \\ 
		Source & Date & $t_{\rm int}$$^a$& $\nu^b$ & $f_{\nu}^c$ & Telescope & Date & $t_{\rm exp}$ & $\fx$$^d$\\
		&   & (min) & (GHz) & ($\mu$Jy) &  &  & (ks) & ($10^{-13}$ erg s$^{-1}$ cm$^{-2}$) \\
		\hline
		\src & 2018 Oct 8--9  & 600 & 7.25 &  $21.7\pm6.4$ &  \textit{Chandra} & 2018 Oct 8 & 19.7 & $0.9 \pm 0.1$  \\
		\src & 2020 Mar 13--14 & 1200  & 7.25 & $23.6 \pm 3.6$ & \textit{Swift} & 2020 Mar 14 & 4.8  & $2.7^{+1.0}_{-1.3}$ \\
		\srctwo & 2018 Oct 9--10& 460 & 5.5 & $<23.4^e$ & ...  & ...  & ...  & ... \\
		\hline
	\end{tabular}

	$^a$ Integration time on source. 
	
	$^b$ Central frequency of radio image.
	
	$^c$ Radio flux density at the central frequency $\nu$.
	
	$^d$ Unabsorbed X-ray flux from 1-10 keV.
	
	$^e$ 3$\sigma_{\rm rms}$ upper limit.
\end{table*}

\subsubsection{\srctwo: 2018 with ATCA}

We observed \srctwo\ from 2018 October 9 UT 18:30 - 2018 October 10 UT 04:30, obtaining $\approx$460 min of integration on source (prior to flagging); we used 0823$-$500 as the bandpass and flux calibrator, and 1133$-$681 as the phase calibrator.  Atmospheric stability and weather were worse for our run on \srctwo, such that we started the  night with 10 min cycles to the phase calibrator, which we shortened to 5 min around UT 01:30 when we started observing through clouds and light rain.  Our observing run was  prematurely ended at UT 04:30 due to thunderstorms and a lightning stow of the antennas (and we ultimately flagged much of the data taken after UT 01:30). 

Data were calibrated in the same manner as described above.  However, our time-dependent phase calibrations were generally of poor quality at 9.0 GHz (worsening toward the end of the run).  At 5.5 GHz the calibrations were of reasonable quality, and we only imaged the lower ($\approx$2 GHz) frequency baseband for flux estimation.  We did not obtain a radio detection to a limit of $f_{5.5\, {\rm GHz}} < 23.4\,\mu$Jy bm$^{-1}$ ($3\sigma_{\rm rms}$).  Even though the observing conditions were less than ideal, the 5.5 GHz observations were of sufficient quality that if ATCA were able to detect radio emission from \srctwo, then we would have expected to see (at a minimum) marginal 1-2$\sigma$ indications of radio emission at the known optical/X-ray position of \srctwo.  Since no signs of radio emission were present, there was no motivation for deeper follow-up ATCA observations on \srctwo.

\subsection{X-ray Observations}

For \src, we obtained quasi-simultaneous X-ray observations during both our 2018 and 2020 observing runs, using  \textit{Chandra} in 2018 and  \textit{Swift} in 2020.

\subsubsection{\textit{Chandra} in 2018}

\textit{Chandra} observations were obtained through Director's Discretionary Time (DDT; PI Reynolds), which started on 2018 October 8 UT 06:03 and lasted for 19.7 ks (obsid 21868).   The target was placed at the aimpoint of the S3 chip on the Advanced CCD Imaging Spectrometer \citep[ACIS;][]{garmire03}, and data were telemetered in VFAINT mode.   Data were reduced using the \textit{Chandra} Interactive Analysis of Observations  v4.12 ({\tt CIAO}; \citealt{fruscione06}) and {\tt CALDB} v4.9.1.  The data were first re-processed with {\tt chandra\_repro}.  We detected 83 counts in a circular aperture with radius 2.5 arcsec, with $\leq$ 1 background counts expected within that aperture (the sky background was measured within an annulus centred on the source position, with inner and outer radii of 4 and 7 arscsec respectively).   

To calculate fluxes, a spectrum was extracted using the {\tt specextract} tool.  The spectrum was then fit using {\tt XSPEC} v12.10.1f, adopting  an absorbed powerlaw model ({\tt tbabs} $\times$ {\tt power}) \citep{wilms00} and W-statistics \citep[see][]{cash79}.  The best-fit column density and photon index were $N_H = \left(9.2 \pm 2.7\right) \times 10^{21}$ cm$^{-2}$ and $\Gamma=1.86 \pm 0.22$.  From these spectral parameters, we calculated an unabsorbed model flux of $\fx = \left(9 \pm 1\right)  \times 10^{-14}$ erg s$^{-1}$ cm$^{-2}$.

\subsubsection{Swift in 2020}
In 2020,  \textit{Swift} DDT observations were taken with the X-ray Telescope (XRT; \citealt{burrows05}) in photon counting (PC) mode (obsID 00033811067; PI Reynolds).  We observed for a total of 4.8 ks, with the observations split over multiple $\approx$1 ks snapshots.  The first snapshot started on 2020 March 14 UT 00:09, and the final one started at UT 09:23.

Data were reprocessed using {\tt HEASoft} v 6.25\footnote{\url{https://heasarc.gsfc.nasa.gov/docs/software/heasoft/}} and the latest version of the \textit{Swift} CALDB.  Data were first reprocessed through {\tt xrtpipeline}.  Source counts were extracted from a circular aperture with radius 10 pixels;  background counts were estimated over an annulus with inner and outer radii of 60 and 90 pixels, respectively.  We  detected \src\ with $\sim$ 12 net counts.

To calculate fluxes, we extracted a spectrum using {\tt xselect}, and we created an ancilliary response file using the tool {\tt xrtmkarf} (and the appropriate response matrix was taken from the CALDB).  We adopted an absorbed powerlaw model in {\tt xspec}, and, given the low number of counts, we fixed the column density to $N_H=9\times 10^{21}$ cm$^{-2}$ (i.e., the best-fit value from our 2018 \textit{Chandra} observation).   Using W-statistics, we found a best-fit photon index of $\Gamma=2.1 \pm 0.6$ and an unabsorbed  model flux of $\fx = 2.7^{+1.0}_{-1.3} \times 10^{-13}$ erg s$^{-1}$ cm$^{-2}$.

\section{Results}
\label{sec:res}

\begin{figure*}
	\includegraphics[width=2.0\columnwidth]{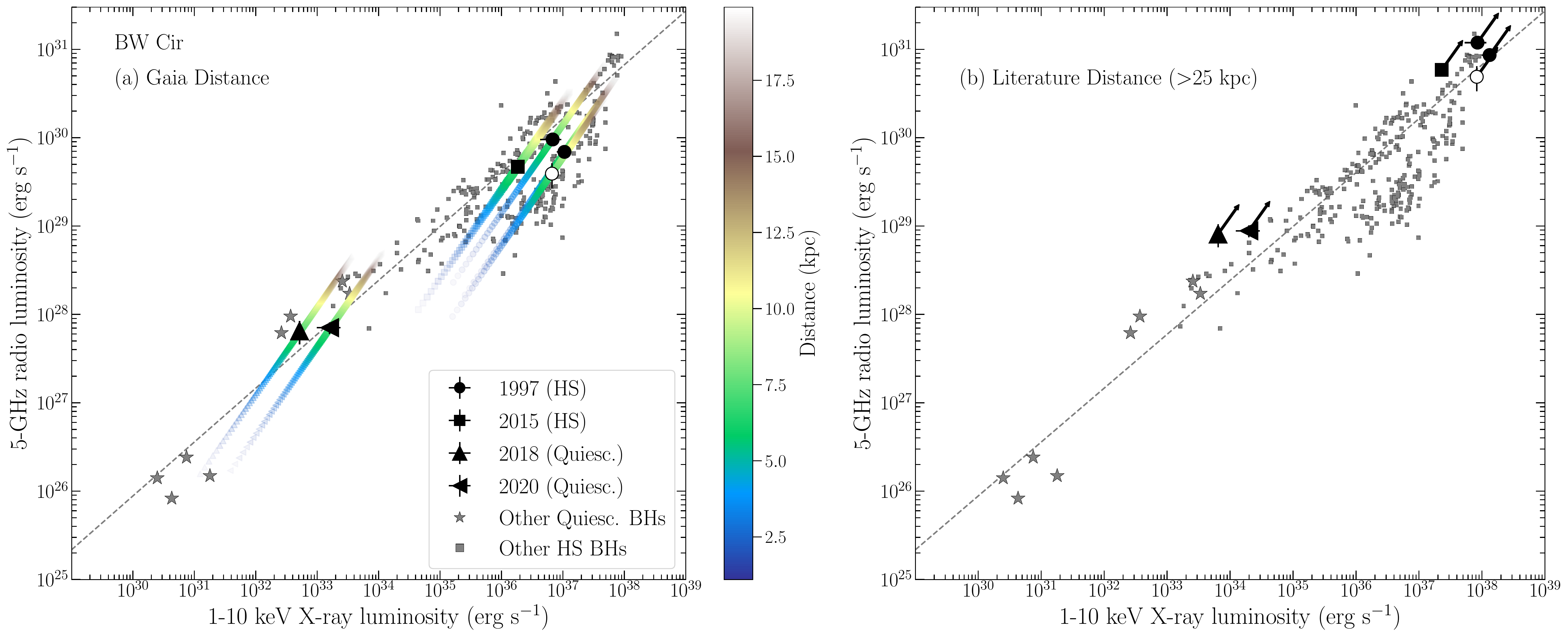}
    \caption{
        Location of \src\ in the radio/X-ray luminosity plane if it is located within the distance range implied by the \textit{Gaia} parallax measurement (panel a) or at $>$25 kpc (panel b; \citetalias{casares09}).  Black/white symbols denote data points in quiescence (in 2018 and 2020) and in the hard state (HS) during its 1997 (\citealt{brocksopp01}) and 2015 (\citealt{coriat15}) outbursts (see legend).  For the HS, black closed symbols indicate that \src\ was either rising out of quiescence or near the outburst peak, and the open symbol denotes the source was decaying back to quiescence.  For panel a, the black symbols represent the mode of the posterior distributions, and  the color  scale represents  luminosities at various distances out to 20 kpc (beyond which the source's space velocity starts to become large; see \S \ref{sec:disc:kindoffar}).   The transparency of each colored data point scales with the value of the posterior.  The grey squares show the population of other hard state \xrb s, taken from the catalog compiled by \citet{bahramian18}.  The grey star symbols denote other quiescent \xrb s with radio detections, including A0620$-$00 \citep{gallo06, dincer18}, XTE J1118+480 \citep{gallo14}, MWC 656 \citep{ribo17}, V404 Cygni \citep{corbel08,hynes09,rana16}, and GX 339$-$4 \citep{corbel13, tremou20}.  Interpretation of this figure depends on the correct distance to \src.}
    \label{fig:lrlx}
\end{figure*}

\subsection{\src}
\label{sec:res:bwcir}
In Figure~\ref{fig:lrlx} we show where \src\ would fall in the radio/X-ray luminosity plane if it is located at the distance(s) implied by the \textit{Gaia} parallax (panel a) vs.\ the $>$25 kpc distance from \citetalias{casares09} (panel b).  
 All radio flux densities are converted from 7.25 to 5.0 GHz by assuming a flat radio spectrum (i.e., $\alpha=0$, where $f_\nu \propto \nu^\alpha$).  To supplement the two data points in quiescence from 2018 and 2020, we searched the literature for other quasi-simultaneous radio and X-ray observations when \src\ was outbursting in the hard state.  We found four additional data points for inclusion on the radio/X-ray plane, three from its 1997 outburst and one from its 2015 outburst.  
 
 During the 1997 outburst, during which \src\ remained entirely in the hard state, \citet{brocksopp01} performed multiwavelength monitoring that included five radio observations with ATCA.  Of those five ATCA observations,   three were taken within $\pm$0.5 d of an X-ray constraint with the All Sky Monitor (ASM) on the \textit{Rossi X-ray Timing Explorer (RXTE).}  We extracted corresponding \textit{RXTE}/ASM count rates from the daily average light curve online data products\footnote{\url{http://xte.mit.edu}} 
and converted to 1-10 keV unabsorbed X-ray fluxes using the online version of the Portable, Interactive Multi-Mission Simulator\footnote{\url{https://heasarc.gsfc.nasa.gov/cgi-bin/Tools/w3pimms/w3pimms.pl}} 
({\tt WebPIMMS}), assuming an absorbed power-law with $N_H = 9 \times 10^{21}$ cm$^{-2}$ and $\Gamma=1.5$.   We thus include  5 GHz radio flux densities\footnote{\citet{brocksopp01} report radio flux densities measured in  both the image plane and in the uv plane.  We adopt the average of their `image' and `uv' flux densities at 4.8 GHz, and we convert to 5 GHz adopting   radio spectral indices presented in their Table~6.}  
and 1-10 keV X-ray fluxes on November 25, December 2, and December 19 of $\fr = 3.2\pm0.3$, $2.3\pm0.3$, and $1.3\pm0.4$ mJy and $\fx= \left(1.1\pm0.4\right)\times10^{-9}$, $\left(1.8\pm0.4\right)\times10^{-9}$, and $\left(1.1\pm0.2\right)\times10^{-9}~\flux$, respectively.  The first epoch (November 25) was taken toward the end of the hard state rise, the second epoch (December 2) near the peak of the outburst, and the final epoch (December 19) toward the beginning of the descent back to quiescence.

During the 2015 outburst, we  found one published  radio observation (which was taken during the hard state rise; \citealt{coriat15}).   ATCA observed on 2015 June 16 UT 06:30-09:30, from which we estimate $f_{5\,{\rm GHz}} = 1.56 \pm 0.02$ mJy (extrapolated to 5 GHz from 5.5 and 9.0 GHz flux densities listed in \citealt{coriat15}, and their spectral index $\alpha= -0.10 \pm 0.05$).  X-ray observations were taken by \textit{Swift} on 2015 June 15 and June 17 (both taken with the XRT in  window timing mode).  We extracted spectra for these two observations using the online \textit{Swift}/XRT data product generator \citep{evans09}, and we fit absorbed powerlaws to both spectra in {\tt XSPEC} (using W-statistics). Unabsorbed model fluxes were calculated for each date and then averaged together to estimate $\fx = \left(3.1 \pm 0.1\right) \times 10^{-10}~\flux$ on 2015 June 16.\footnote{The best-fit spectral parameters are consistent on each date.  On 2015 June 15 ($\approx$3600 counts) we find $N_H=7.0_{-0.7}^{+0.4} \times 10^{21}$ cm$^{-2}$, $\Gamma=1.29\pm0.04$, and $\fx=\left(2.9\pm0.1\right) \times 10^{-10}$ erg s$^{-1}$ cm$^{-2}$.  On 2015 June 17 ($\approx$3000 counts) we find $N_H=\left(7.0\pm0.5\right) \times 10^{21}$ cm$^{-2}$, $\Gamma=1.35\pm0.04$, and $\fx=\left(3.2\pm0.1\right) \times 10^{-10}$ erg s$^{-1}$ cm$^{-2}$.}  

\subsection{\srctwo}
\label{sec:res:srctwo}
\begin{figure}
 	\includegraphics[width=\columnwidth]{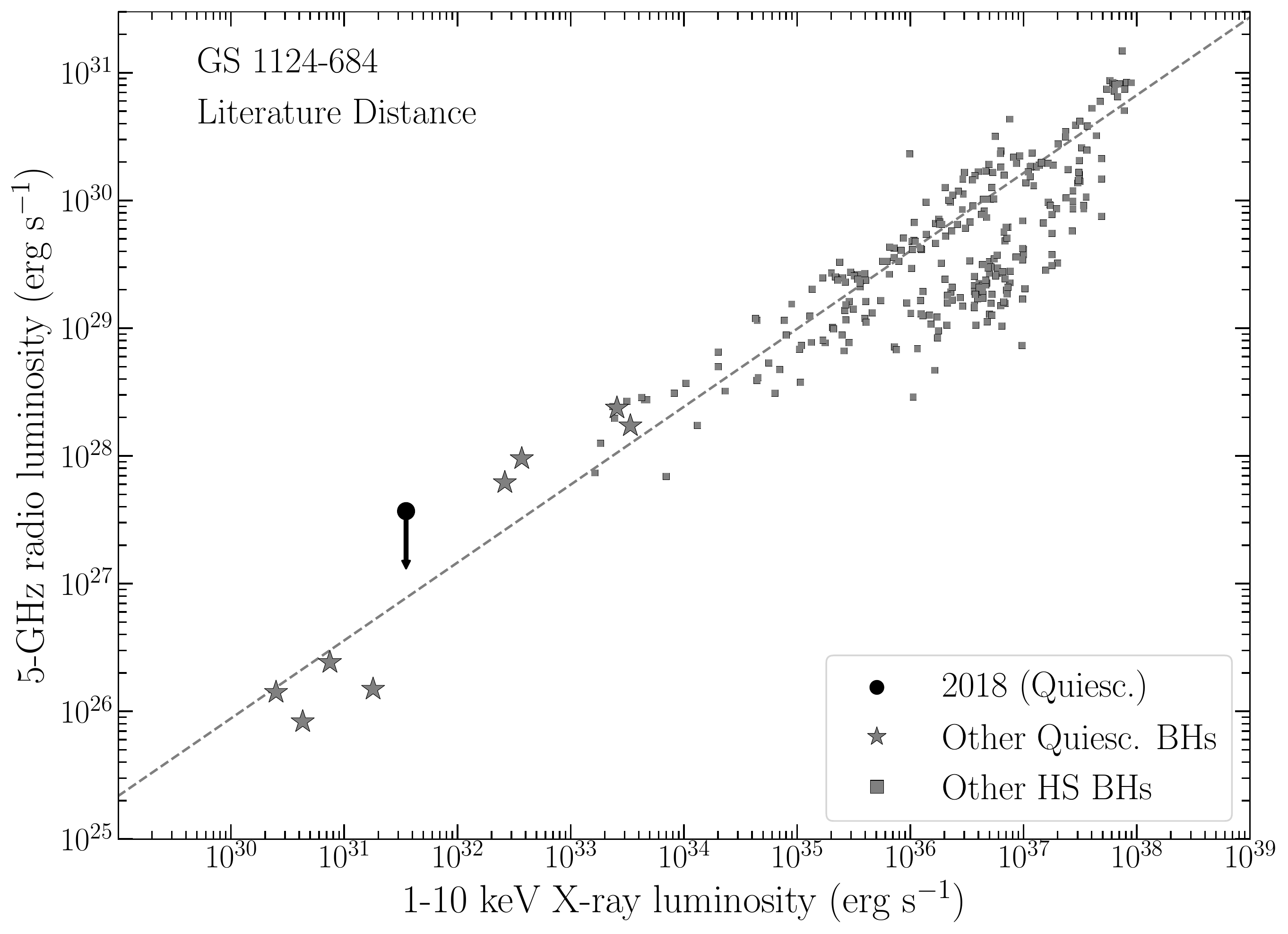}
    \caption{ Limits on the location of \srctwo\ in the radio/X-ray luminosity plane, adopting the literature distance of $4.95^{+0.69}_{-0.65}$ kpc (\citealt{wu16}).   All other symbols have the same meaning as in Figure~\ref{fig:lrlx}. Our limit on the radio luminosity does not place strong constraints on the presence or absence of a radio jet from \srctwo.}
    \label{fig:gs1124_lrlx}
\end{figure}

In Figure \ref{fig:gs1124_lrlx} we show limits for \srctwo\ in the radio/X-ray luminosity plane.  Since the radio non-detection is not meaningful at any distance implied by \textit{Gaia}, we illustrate its location in Figure~\ref{fig:gs1124_lrlx} by only using the more precise donor-star based distance from \citealt{wu16}.  \srctwo\ has only had one X-ray observation in quiescence, by \textit{XMM-Newton} in 2001 \citep{sutaria02}.  Since we did not coordinate X-ray observations with our 2018 ATCA observation, we use  the 2001 X-ray information for placing \srctwo\ on $\lr - \lx$, despite the 17 year time difference.  Using the best-fit spectral parameters ($\nh = 2.6 \times 10^{21}$ cm$^{-2}$; $\Gamma=1.6\pm0.7$) and absorbed X-ray flux ($1.3\times10^{-14}~\flux$ from 0.3-12 keV) reported by  \citet{sutaria02}, who detect \srctwo\ with $\approx$100 counts,  we adopt an unabsorbed X-ray flux of $\fx = 1.2\times10^{-14}~\flux$ (we ignore errors on flux measurements since uncertainties  are dominated by  non-simultaneity).  Unfortunately, the new ATCA radio limit is not constraining in terms of the disk/jet coupling in \srctwo.  Also, there are not any archival radio observations of \srctwo\ in the hard state from its 1991 outburst \citep[see \S 3.4 of][]{fender01}. Thus, we do not discuss \srctwo\ further in this paper, other than to say our radio non-detection is consistent with expectations given the $4.95^{+0.69}_{-0.65}$ kpc distance from \citet{wu16}.

\section{Discussion}
\label{sec:disc}

\src\ now represents the sixth  \xrb\ with a radio detection in quiescence (see the star symbols in Figure~\ref{fig:lrlx}; note three systems appear twice in the Figure). The correct physical interpretation of \src's behaviour in $\lr-\lx$  depends on the correct distance, and from the radio/X-ray information alone we cannot make strong conclusions. So, in the following discussion  we outline implications on the disk/jet coupling of \src\ within different distance ranges.  Although we discuss all possible distances  for completeness, we stress that  any distance $\lesssim 13$ kpc poses a severe challenge in that it would require a  small donor  mass  ($M_d \lesssim 0.1~\msun$; see \S \ref{sec:targs:bwcir}) since the mass   decreases rapidly with distance ($M_d \propto d^3$; assuming that the donor fills its Roche Lobe).
Throughout, calculations on the `quiescent X-ray luminosity' are derived from the minimum X-ray flux (1-10 keV) that we observed:  $8.6_{-1.0}^{+1.4} \times 10^{-14}$ erg s$^{-1}$ cm$^{-2}$ in 2018, which we note is similar to the flux obtained during a  2010 \textit{Chandra} observation in quiescence ($\fx = 9.1^{+0.7}_{-0.5} \times 10^{-14}~\flux$ from 1-10 keV; \citealt{reynolds11}).

\subsection{Scenario 1: $\mathbf{d \lesssim 5}$ kpc}
\label{sec:disc:verynearby}
All hard state and quiescent \xrb s were once thought to follow a correlation of the approximate form $\lr \propto \lx^{0.6}$, which has colloquially been referred to as the `standard'  track in the radio/X-ray luminosity plane \citep[e.g.,][]{coriat11, gallo14}.  However, it now appears that the majority of hard state \xrb s are in fact `radio-faint', meaning that they emit less radio emission than expected in the hard state, at least when $\lx \gtrsim 10^{36}~\ergs$ \citep[e.g.,][]{corbel04, cadolle-bel07, coriat11, meyer-hofmeister14, tomsick15, espinasse18, motta18}.   All `radio-faint' systems that have so far been monitored with sufficient radio sensitivities  appear to start moving horizontally in the $\lr-\lx$ plane as $\lx$ decreases below $\approx10^{36}~\ergs$.  Those systems then eventually  rejoin the `standard' track around $\lx \approx 10^{34}-10^{35}~\ergs$ \citep{coriat11, plotkin17a}.

If \src\ were to reside  very nearby at the smallest distances allowed by the \textit{Gaia} parallax measurement ($d \lesssim 2$ kpc; a 4\% chance from integrating the EDR3 posterior),  then it would mark the only \xrb\ observed to continue down the `radio-faint' track when $\lx \lesssim 10^{36}~\ergs$.  In this case though, it still does not appear that the `radio-faint' track extends indefinitely into quiescence, as our quiescent radio detections lie above the extrapolation of the `radio-faint' track at the lowest X-ray luminosities (unless the quiescent radio emission is dominated by the companion star; see Section~\ref{sec:disc:corona}).    At this distance,  the quiescent X-ray luminosity would be
$\lesssim 4 \times 10^{31}~\ergs$, which is lower than expected given the system's orbital period.  
   A small distance would then  imply an accretion flow that is even more radiatively inefficient and underluminous than other quiescent systems.  
  However, we can rule out this scenario because a distance $<$2 kpc would require an implausibly low-mass companion star ($M_d \lesssim 4 \times 10^{-4} \msun$). 

Extending the distance slightly upward to $\approx$2--5 kpc (18\% chance) still requires an incredibly low-mass donor   ($M_d \lesssim 6 \times 10^{-3} \msun$), but it would yield X-ray luminosities consistent with expectations for the orbital period ($\lx \approx 4\times 10^{31} - 3 \times 10^{32}~\ergs$; \citealt{reynolds11}).   In fact, this distance range provides the  only scenario that would allow \src\ to display what is currently viewed as `normal' behaviour.   For example, in quiescence it would fall on the $\lr - \lx$ `standard' track to the lower-left of V404 Cygni,  as expected given the 2.5 vs.\ 6.5 d orbital periods of \src\ vs. V404 Cygni \citep{casares92}.   \src\ then affords an exciting laboratory for accretion flow and jet modeling, as it would be the first radio-detected quiescent \xrb\ to start  filling in the (current) gap in $\lr-\lx$ between V404 Cygni (at $\lx \approx 10^{33}~\ergs$; \citealt{bernardini14}) and the other radio-detected quiescent \xrb s at $\lx \approx 10^{31}~\ergs$ (A0620$-$00, XTE J1118+480, and MWC 656; \citealt{gallo06, gallo14, ribo17, dincer18}).

Furthermore, at a distance between 2--5 kpc,  \src\ would  be classified as a `radio-faint' hard state \xrb\ during outburst, and it would mark the first quiescent radio detection of a (known) `radio-faint' \xrb.\footnote{It is unknown if A0620$-$00 was  `radio-faint' during outburst.}  
The combined  quiescent radio and X-ray emission would then suggest that \src\ transitioned back to the `standard' radio/X-ray correlation as it moved from the hard state back to quiescence.  Noting that   all of our hard state data points in Figure~\ref{fig:lrlx} lie on the `radio-faint' track, and two of our data points were taken during the outburst rise (one in 1997 and one in 2015),  then this scenario would provide new empirical evidence that (at least some) `radio-faint' \xrb s rise out of quiescence through the `radio-faint' track (opposed to a `hysteresis' effect where they could rise out of quiescence along the `standard' track, and then fade back to quiescence along the `radio-faint' track). 

\subsection{Scenario 2: $\mathbf{5 \lesssim d \lesssim 20}$ kpc}
\label{sec:disc:kindoffar}

If \src\ were to fall at the higher end of the distance range allowed by \textit{Gaia}, then it would still fall on the `standard' track in quiescence.  Note, the 1$\sigma$ \textit{Gaia} distance range extends up to $\approx$12 kpc, but the posterior distribution allows distances up to $\approx$20 kpc or slightly higher (see Figure~\ref{fig:dist}).  There is a 76\% chance that \src\ falls between 5-20 kpc,
(calculated by integrating the EDR3 posterior), and only a 2\% chance the distance is $>$20 kpc.    Portions of the 5-20 kpc distance range would  place \src\ within spiral arms of the Milky Way: the Scutum-Centaurus arm if  $5\lesssim d \lesssim10$ kpc and the Sagittarius arm if $d\approx15$ kpc.  Intriguingly, distances up to $\approx$15 kpc
imply that the 1997 and 2015 data points during the outburst rise follow the `standard' track, and the data points during the 1997 outburst peak and decay fall  on the `radio-faint' track.  Such a scenario would imply that \xrb s can show a hysteresis between the hard state rise and the decay \citep[e.g.,][]{russell07, islam18}.\footnote{At $d \approx15$ kpc, the switch between tracks would occur near $3 \times 10^{37}~\ergs$, which is around  the luminosity where the `standard' and `radio-faint' tracks start to join each other.}  
A distance of 10--15 kpc would also start bringing the peak outburst luminosity in line with expectations (i.e., $\approx$10\% of the Eddington luminosity, if one adopts a bolometric correction of $\approx$10); however, then the quiescent X-ray luminosity would  be between $3\times 10^{32} - 10^{33}~\ergs$ and therefore above expectations from the quiescent $\lx-P_{\rm orb}$ relation.

If the distance is $\gtrsim$15 kpc, then the source would always fall along the standard track, albeit at higher X-ray luminosities than expected given its orbital period (both in quiescence and at peak outburst).  Nevertheless, there is not an obvious argument in terms of its radio/X-ray luminosity coupling to disfavour such a scenario.  Note, distances between 16--20 kpc are formally consistent with both the \textit{Gaia} posterior distribution and the donor mass-based distance (given uncertainties in the reddening toward \src; \citealt{reynolds11}).

\begin{figure}
 	\includegraphics[width=\columnwidth]{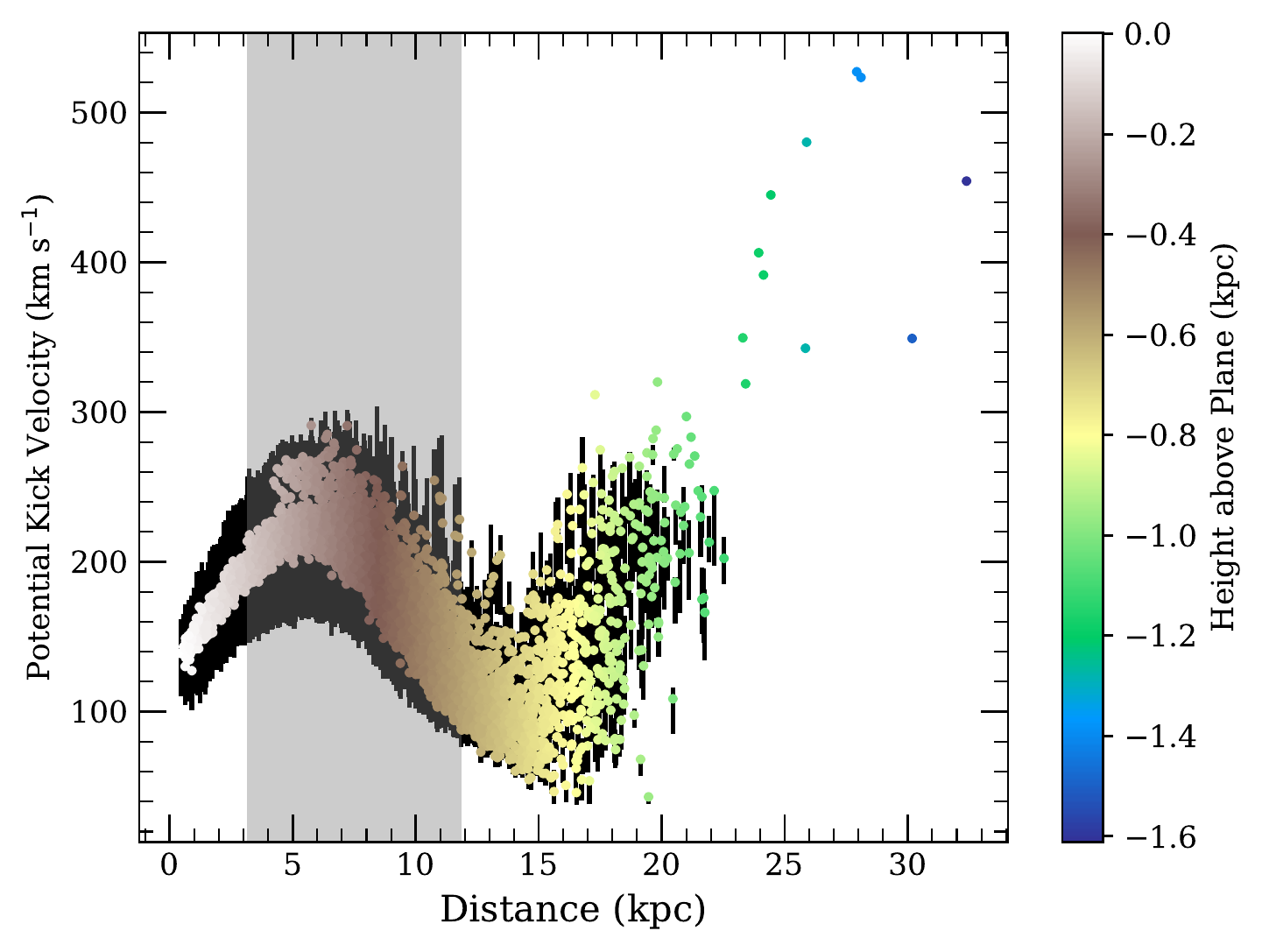}
    \caption{Variation of potential kick velocity (PKV) with distance. The error bars (shown in black) represent the 5$^{\mathrm{th}}$ and 95$^{\mathrm{th}}$ percentiles of the PKV distribution for a given distance. The colourbar represents the height of the source from the Galactic Plane for a given distance. The gray shaded area is the 68$\%$ confidence interval of the distance posterior derived using the \textit{Gaia} EDR3 parallax measurement. The PKV stays $\lesssim 300$ km s$^{-1}$ for up to $\approx$20 kpc. The stark decrease in the number of data points above $\approx$20 kpc (and the lack of error bars) is a combined effect of the shape of the distance posterior distribution that was used as input to the simulations (see Figure~\ref{fig:dist}), and that only for few realisations does the  source  cross the Galactic plane  at those distances. }
    \label{fig:pkv}
\end{figure}

Based on the \textit{Gaia} DR2 proper motion and parallax measurements, \citet{gandhi19} argued that distances $\gtrsim 10$ kpc are unlikely because then \src\ would start to have an excessively large space velocity.  However, the systematically lower proper motion and parallax measurements in EDR3 (see Table~\ref{tab:gaia}) make distances even as large as $\approx$20 kpc more comfortable, as described below. We estimate the potential kick velocity (PKV) that the system would need for it to move with the measured proper motion and systemic radial velocity, assuming the system was born in the Galactic plane. We used the methodology developed by \citet{atri19} that uses Monte Carlo simulations to trace back the Galactocentric orbit of the source.  We ran 5000 instances to estimate the distribution of the velocity of the system when it crosses the Galactic plane (i.e., at $z=0$). Figure~\ref{fig:pkv} shows the PKV of \src\ as a function of distance and height above the Galactic plane, where the distance is sampled from the distance posterior distribution using the EDR3 parallax, and the simulations are based on the \textit{Gaia} EDR3 proper motions and a radial velocity of $103 \pm 4$ km s$^{-1}$ \citepalias{casares04}. Note, we adopt the PKV as a tracer of the motion of the source instead of peculiar velocity (i.e., the space velocity compared to the local standard of rest) because the local neighborhood of the source becomes poorly defined as  the height from the Galactic plane increases.

The PKV distribution has a mode and 68\% confidence interval of $165^{+81}_{-17}$ km s$^{-1}$. More importantly, we note that the PKV of \src\ stays within reasonable values of $\lesssim 300$ km s$^{-1}$  out to $\approx$20 kpc (such velocities have been observed in other \xrb s; \citealt{atri19}). This then suggests that based on the EDR3 measurements, the PKV of \src\ up to a distance of $\lesssim 20$ kpc is feasible.   Excluding larger distances based on the space velocity only removes $\approx$2\% of solutions formally allowed by the  distance posterior from the EDR3 parallax measurement.

\subsection{Scenario 3: $d>25$ kpc}
\label{sec:disc:veryfar}

If \src\ is at $>$25 kpc (as implied by the properties of the donor star), then we reach similar conclusions as for distances 15--20 kpc in the previous subsection: \src\ always falls along the `standard' track of the radio/X-ray luminosity correlation, and it would be  in an unusually luminous accretion state in quiescence (relative to its orbital period; \citealt{reynolds11}).  However, a distance $>$25 kpc represents only 1\% of solutions allowed by the \textit{Gaia} EDR3 distance posterior, and at these distances the space velocity may become unphysical if the system is bound to the Galaxy (Figure~\ref{fig:pkv}).  Furthermore, $d>25$ kpc would  place \src\ at a higher elevation above the Galactic plane than expected for its orbital period \citep{gandhi20}.   Note, if the X-ray luminosity reached by GX 339$-$4 between outbursts is truly its lowest quiescent luminosity ($\approx 3\times10^{33}~\ergs$ at 8 kpc; \citealt{tremou20}), then that source would also be in an elevated quiescent accretion state relative to its $\approx$42 h orbital period \citep{hynes03, heida17},  giving additional credence to the idea of an unexpectedly luminous quiescent accretion state for \src.

\subsection{On radio emission from the donor star}
\label{sec:disc:corona}
As detailed below,  the only way that our quiescent \src\ radio detection  could originate  from the donor star instead of a relativistic jet is if \src\ happens to fall in the low-distance tail of the \textit{Gaia} posterior distributions.   
 Stellar coronal emission is usually gyrosynchrotron radiation, the luminosity of which  is known to correlate with the X-ray luminosity of active stars (i.e., the G{\"u}del-Benz relation; \citealt{guedel93, benz94}).  The most luminous radio/X-ray stars reach $\lr \approx 5\times 10^{27}~\ergs$ and $\lx \approx 7\times10^{31} \ergs$\ (we converted luminosities on the G{\"u}del-Benz relation to our 5 GHz and 1-10 keV bandpasses assuming a flat radio spectrum and an X-ray photon index of 2).  These  luminosity extremes are displayed by a combination of RS CVn and FK Comae systems.  Some of these  make decent comparisons  to \src,  since RS CVn stars are binary systems where one of the components is a G-K giant/sub-giant.  FK Comae stars are single stars, but with a similar spectral type (G-K giant/sub-giant) and rotation speed as the secondary in \src: $v \sin i \sim 100$ km s$^{-1}$ for FK Comae stars \citep{bopp81} vs.\ $v\sin i  \approx 70$ km s$^{-1}$ for \src\ \citepalias{casares09}.

Considering the above, and scaling from our 2018  observations of \src\ ($f_\nu = 21.7 \mu$Jy and $\fx=9\times10^{-14}~\flux$), the most extreme stellar systems are consistent with the quiescent $\lx$\ of \src\ only out to $\approx$2.6 kpc.\footnote{The chance that we observed  a radio flare is very low, especially since we measured similar radio flux densities on two different occasions.}  
However, at that distance the observed radio emission from \src\ would be 4 times larger than expected from the G{\"u}del-Benz relation, and that is assuming that all of the X-ray emission is emitted by the donor star and not the accretion flow (which is extremely unlikely).  Thus, 2.6 kpc is a  generous upper limit on donor star radio emission.  A more realistic estimate is obtained by considering that the average radio luminosity of radio-detected RS CVn systems (with periods similar to \src) is $\lr \approx 10^{26}~\ergs$ \citep{drake89}, which would only be detectable out to $\approx$0.7 kpc.

\section{Future work: toward reconciling the \textit{Gaia} and Casares et al.\ distances}
\label{sec:disc:future}

From the radio and X-ray properties alone, we cannot strongly favour or disfavour any particular distance. That said, given the system's current mass function and donor spectral type (including the detection of ellipsoidal modulations implying a Roche-lobe fed system; \citetalias{casares04, casares09}), it is extremely difficult to find a way to make the binary system fit at a distance $\lesssim$13 kpc.  For a distance $2\lesssim d \lesssim 13$ kpc to be correct, then the distance modulus must decrease by  1.4 to 5.5 mag (relative to the distance modulus of 17 mag calculated by \citetalias{casares09}). This requires a donor star luminosity 4--160 times smaller, and/or corrections to the optical apparent magnitude  must be more extreme than currently accounted for  (i.e., due to disk veiling, extinction, etc.).  

While  tweaks are possible to the donor star luminosity, it is difficult to imagine how the luminosity calculated by \citetalias{casares09} could be off by upwards of two orders of magnitude. Simply requiring a lower-mass donor  is not a good solution, since the mass function and $q$ value limit the donor mass to $\gtrsim0.9 \msun$.  We therefore only envision two areas of future investigation that might be worthwhile.  The first would be to repeat the \citetalias{casares04, casares09} analyses  in multiple  filters (i.e., their distance modulus was estimated only in the Bessel $R$ filter), although we are doubtful that would provide a meaningful revision.  The other area that might show more promise (but is out of the scope of this paper) would be to explore  model atmospheres for G sub-giants in close binaries.  Perhaps there is an additional opacity source or unusual stellar structure that could cause a lower luminosity (compared to the current estimate based on the Stefan-Boltzmann Law followed by  bolometric and color corrections into the $R$ filter).  

Regarding potential apparent magnitude corrections, significant adjustments to the disk veiling seem  unlikely (see the discussion in \citetalias{gandhi19}).  The largest source of uncertainty is  probably in the  reddening.  The $E\left(B-V\right)\sim 1$ value adopted by \citetalias{casares09}  was based on the optical colors of \src\ during its 1987 outburst \citep{kitamoto90}, and also from the equivalent widths of interstellar absorption lines \citepalias{casares04}.   As noted by \citet{reynolds11},  the interstellar absorption lines could imply $E\left(B-V\right)$ as large as $\approx$1.4, which they discuss is also in line with expectations from the amount of X-ray absorption and would imply a distance $\gtrsim$16 kpc. Increasing the reddening would then allow a narrow range of $16\lesssim d \lesssim 20$ kpc where the \textit{Gaia} EDR3  and \citetalias{casares09} distances are consistent. 
Note, however, that \citet{koljonen16} place a more restrictive  limit on  $E\left(B-V\right)$.   They suggest that the extinction coefficient $A_V$ cannot be larger then 3.5 mag (or equivalently $E\left(B-V\right) < 1.1$), or else the optical/ultraviolet spectral energy distribution would be unphysically steep.  For $A_R/E\left(B-V\right)=2.6$ \citep{schlegel98}, a reddening of 1.1 would only lower the distance modulus by 0.3 mag (i.e., $d\gtrsim 22$ kpc), which then  excludes the possibility of reconciling  \citetalias{casares09} with current \textit{Gaia} measurements.

Of course, an alternative solution  is that the  \textit{Gaia} parallax is in error.  Ultimately, direct verification is going to require a more precise parallax measurement, which  might become available in future \textit{Gaia} data releases, or, more likely, could  be achieved in the radio waveband via very long baseline interferometry (VLBI).   So far, four \xrb s have had parallaxes measured from radio VLBI observations \citep{miller-jones09, reid11, reid14, atri20}.  Unfortunately, the $\approx$20-25 $\mu$Jy quiescent radio flux density of \src\  precludes a VLBI radio parallax using current facilities in quiescence, and it is unclear if it would be in reach with future (near-term) facilities in the southern sky.  
Given that \src\ reaches mJy flux levels during outbursts \citep{brocksopp01, coriat15}, a radio parallax could, however, be measured with current telescopes via appropriately timed  (hard state) observations during a future outburst \citep[e.g.,][]{atri19}.  

\subsection{Source Confusion in Gaia?}
\label{sec:disc:gaiaconfusion}
In the near-term, we believe the most promising distance investigations should focus on if the \textit{Gaia} astrometry is in error because of a faint interloper blending with the optical counterpart.  Given that the \textit{Gaia} source identification is very likely correct (\citetalias{gandhi19}), and quality flags for \src\ in the \textit{Gaia} database all suggest a high-quality detection of a single source,  this idea is likely only plausible if the culprit source  contributes  $<$10-20\% of the quiescent optical flux.\footnote{The  \textit{Gaia} quality flags {\tt astrometric\_gof\_al = 1.8} and {\tt ruwe = 1.1 } in EDR3, which should nominally be $\lesssim$3 and $\approx$1, respectively, for good astrometric fits using a single star model.  See \url{https://gea.esac.esa.int/archive/documentation/GEDR3/Gaia_archive/chap_datamodel/sec_dm_main_tables/ssec_dm_gaia_source.html}} 
Note, 10--20\% optical flux contamination would not affect the \citetalias{casares09} distance in any meaningful way.  However, if an interloper were any brighter, then \citetalias{casares09} would have very unlikely detected ellipsoidal modulations.   For this to be a viable solution, the interloper should be closer than \src\ to bias the astrometric motions to higher values.  As described below, the probability is low but not negligible. 

In EDR3, the completeness of resolved pairs of close stars drops rapidly for pair separations $\lesssim$0.7 arcec, and during source transits \textit{Gaia} does not always resolve close pairs of stars with separations $\lesssim0.2-0.3$ arcsec.\footnote{\url{https://gea.esac.esa.int/archive/documentation/GEDR3/index.html}}  
  Thus, in the following we consider that an interloper could be aligned within $\approx$0.2--0.7 arcsec of the \textit{Gaia} position for \src, which corresponds to  a solid angle of $\Omega = 3\times10^{-12} - 4\times10^{-11}$ steradians.    
   In a cone of that solid angle, the total volume is $\frac{\Omega_{\rm star}}{3}d_h^3$, where $d_h$ is the distance out to which the line of sight intercepts a scale height of the Galactic plane.  For the line of sight toward \src, this will be at about 2~kpc, giving a total volume of about 0.008-0.1 pc$^3$.  Taking the local space density of stars to be about 0.1 pc$^{-3}$, this yields a 0.08-1\% chance of the line of sight intercepting another star.    While this is not particularly high, it could have happened with any of the X-ray binaries with \textit{Gaia} parallax measurements, and hence there is a reasonable chance, $\approx$3--30\%, that at least one X-ray binary will have an interloper.\footnote{The 3--30\% chance is according to a binomial probability distribution, assuming there are $\approx$40 neutron and black hole X-ray binaries with \textit{Gaia} parallax measurements. Even just the 11 \xrb s with \textit{Gaia} DR2 parallax measurements in \citetalias{gandhi19} have a 1--10\% chance of including at least one interloping source.}  

Furthermore, not included in the above estimate is that the outskirts of at least two nearby open star clusters cover the line of sight toward \src: Platais 11 ($d=232$ pc) and Platais 12 ($d=402$ pc), also known as  HIP 67330 and HIP 67740 \citep{platais98}.   These clusters have projected radii of 1.5 and 1.0 deg, and \src\ lies 1.7 and 1.5 deg from their centres, respectively.  More generally, this area of the sky is also near the Sco OB2 association, making the probability of a foreground interloping source even higher.

It is  worth reiterating our suspicions that the EDR3 astrometric measurements may be displaying \textit{systematic} changes from DR2 (see \S \ref{sec:targs:bwcir}).  In particular, the distance posterior distribution is now unimodal instead of bimodal (Figure~\ref{fig:dist}).  Additionally, the DR2 proper motion ($\mu_\alpha \cos \delta = -9.38\pm2.22$ km s$^{-1}$, $\mu_\delta = -5.70\pm2.26$ km s$^{-1}$) was in the same range as the median proper motions of the stars in both Platais 11 ($\mu_\alpha \cos \delta = -13.22\pm0.42$ km s$^{-1}$, $\mu_\delta = -8.71\pm0.45$ km s$^{-1}$) and Platais 12 ($\mu_\alpha \cos \delta = -8.29\pm0.28$ km s$^{-1}$, $\mu_\delta = -5.04\pm0.35$ km s$^{-1}$). Such behaviour would be expected if the interloper is in one of these clusters, and if it happened to bias  the astrometric solutions more in DR2  than in EDR3.  

 Naturally the obvious test is to continue to monitor the proper motion of \src\ to see if it continues to  decrease over time (which would allow for higher distances to \src). In the meantime, evidence of an interloper might already be accessible via near-infrared (NIR) or infrared (IR) observations.  If we assume that the interloper contributes 10\% of the total flux measured by \textit{Gaia} ($G=20.6$), then the interloping star has $G=23.2$.  At the distance of Platais 11 and Platais 12, the interloper would have an absolute magnitude of $M_G = 16.4$ and 15.2, respectively, making it an M or L type dwarf star or brown dwarf (see Figure 9 of \citealt{gaiahrdiagrams}).  In the NIR, such a star would have an absolute magnitude in the $K_s$ filter from 10-12, or an apparent magnitude $K_s \approx 17-19$ in Platais 11 or $K_s \approx 18-20$ in Platais 12.

 The NIR magnitude of the quiescent counterpart to \src\ is unknown, except that it is not detected by the Two Micron All Sky Survey \citep[2MASS;][]{skrutskie06} such that $K_s \gtrsim 14$.  There is an IR counterpart in the catalog of `unblurred coadds of the \textit{WISE}\footnote{The \textit{Wide-field Infrared Survey Explorer} \citep{wright10}.}  imaging database' (unWISE; \citealt{lang14, schlafly19})  
that is 0.8 arcsec from the \textit{Gaia} position.  This IR source has flux densities in the W1 (effective wavelength $\lambda_{\rm eff}=3.35\,\mu$m) and W2 ($\lambda_{\rm eff}=4.6\,\mu$m) filters of $f_{3.35}=468 \pm 6$ and $f_{4.6}=372 \pm 6$ $\mu$Jy, respectively.  Upon visual inspection, the coadded unWISE images are of high quality without any obvious artifacts near the location of \src.  Furthermore, the \textit{WISE} data were taken in 2010, when \src\ was not in an outburst, such that $\approx$400 $\mu$Jy appears to be the IR quiescent level. Thus, if we assume that the $K_s$ flux density is somewhere between the quiescent radio measurement ($\approx$20$\mu$Jy) and the quiescent IR measurement ($\approx$400$\mu$Jy), then \src\ should have $16 \lesssim K_s \lesssim 19$ in quiescence \citep[adopting a $K_s$  zero point of 666.8 Jy;][]{cohen03}.  The \src\ magnitude may then be comparable to the NIR magnitude of an interloping star if it resides in Platais 11 or Platais 12.  Since one stands a better chance of spatially resolving an interloper star from \src\ if the two sources have similar fluxes, the NIR (and possibly IR) offers a better opportunity than the optical.  Such an experiment would be feasible at these magnitudes with, e.g., the \textit{Hubble Space Telescope} or the \textit{James Webb Space Telescope.}  Or, if \src\ and an interloper have magnitudes toward  the brighter end of our estimated  range, with, adaptive optics on ground-based telescopes (e.g.,  Gemini-South).  In all cases,  spatial resolutions $\lesssim0.1$ arcsec are achievable.

\section{Summary}
\label{sec:summary}

 \textit{Gaia}   motivated us to search for radio jets from two quiescent \xrb s with  parallax measurements.  One source, \srctwo, was not detected by ATCA.  For the second source, \src, a radio outflow was detected on two separate epochs.  We also obtained  quasi-simultaneous X-ray detections with \textit{Chandra} and \textit{Swift}, respectively, allowing us  to place \src\ on the radio/X-ray luminosity plane.  After also considering archival radio and X-ray information during previous outbursts, we discuss implications for the disk/jet coupling of \src\ depending on if the source is located $\lesssim$20 kpc away (as implied by the \textit{Gaia} EDR3 parallax and proper motion measurements), or if it is $
\gtrsim$25 kpc away (based on the donor star).  The phenomenological behaviour of \src\ in the $\lr-\lx$ plane varies  depending on the correct distance, ranging from \src\ always being a `radio-faint' \xrb\ in the hard state (if $d\lesssim 5$ kpc), to \src\ showing a hysteresis between being a `standard' and `radio-faint' \xrb\ during its hard state rise and decay (if $5 \lesssim d \lesssim 15$ kpc), to \src\ always being a `standard' track \xrb\ (if $d\gtrsim 15$ kpc). 

At the moment, reconciling the low- and high-distances implied by \textit{Gaia} vs.\ its donor star is a  challenge.  We find only a narrow distance range, $16 \lesssim d \lesssim 20$ kpc, that is (a) consistent with the \textit{Gaia} measurements and the donor star (if one allows for slightly more interstellar reddening than assumed by \citetalias{casares09}), and that (b) also allows a reasonable donor mass and space velocity for the system.  If \src\ does not fall in this 16--20 kpc distance range, then potential solutions could be that \src\ has an unusual atmospheric structure causing it to be less luminous than a typical G subgiant, or that the \textit{Gaia} parallax and proper motions are in error possibly because of a faint interloper blending with the optical counterpart in \textit{Gaia}.  It may be possible to resolve such an interloper from \src\ in the NIR or IR with
ground-based adaptive optics and/or space-based imaging. Otherwise, given its faint radio flux density in quiescence, the best opportunity for a more precise parallax measurement will likely need to await  VLBI observations during a future hard state outburst.

\section*{Acknowledgements}
We thank the referee for helpful comments that improved this manuscript.  We thank John Paice for collating measurements from the \textit{Gaia} EDR3 archive.   We are grateful to the \textit{Chandra X-ray Observatory} and to the \textit{Neil Gehrels Swift Observatory} for granting our requests for Director's Discretionary Time.  JCAM-J is the recipient of an Australian Research Council Future Fellowship (FT140101082), funded by the Australian government.  PG thanks STFC for support. The Australia Telescope Compact Array is part of the Australia Telescope National Facility which is funded by the Australian Government for operation as a National Facility managed by CSIRO.  We acknowledge the Gomeroi people as the traditional owners of the Observatory site.  The scientific results reported in this article are based in part on observations made by the \textit{Chandra X-ray Observatory}, and this research has made use of software provided by the Chandra X-ray Center (CXC) in the application package {\tt ciao}. This work made use of data supplied by the UK \textit{Swift} Science Data Centre at the University of Leicester. This research includes results provided by the ASM/\textit{RXTE} teams at MIT and at the \textit{RXTE} SOF and GOF at NASA's GSFC
This research has made  use  of  data  and  software  provided  by  the  High Energy  Astrophysics  Science  Archive  Research  Center (HEASARC), which is a service of the Astrophysics Science Division at NASA/GSFC.  This work has made use of data from the European Space Agency (ESA) mission
{\it Gaia} (\url{https://www.cosmos.esa.int/gaia}), processed by the {\it Gaia}
Data Processing and Analysis Consortium (DPAC,
\url{https://www.cosmos.esa.int/web/gaia/dpac/consortium}). Funding for the DPAC
has been provided by national institutions, in particular the institutions
participating in the {\it Gaia} Multilateral Agreement.  This research made use of Astropy,\footnote{http://www.astropy.org} a community-developed core Python package for Astronomy \citep{astropy-collaboration13, astropy-collaboration18}.

\section*{Data availability}
The data underlying this article were accessed from the Australia Telescope Online Archive (\url{https://atoa.atnf.csiro.au}), from the \textit{Chandra} online archive (\url{https://cda.harvard.edu/chaser/}), and from the High Energy Astrophysics Science Archive Research Center (\url{https://heasarc.gsfc.nasa.gov}).  The derived data generated in this research will be shared on reasonable request to the corresponding author.




\bibliographystyle{mnras}







\bsp	
\label{lastpage}
\end{document}